\documentclass[twocolumn,prl]{revtex4}
  \usepackage[dvips]{graphicx}
\usepackage{dcolumn}
\usepackage{amsmath}
\usepackage{latexsym}

\begin{document}

\title[Short Title]{Superconducting phase qubit based on the Josephson oscillator with strong anharmonicity}
\author{A. B. Zorin$^1$ and F. Chiarello$^2$}
%\affiliation{Affiliation 1}
%\affiliation{Affiliation 2}

\affiliation{$^1$Physikalisch-Technische Bundesanstalt, Bundesallee
100, D-38116 Braunschweig, Germany}

\affiliation{$^2$Istituto di Fotonica e Nanotecnologie, CNR, via
Cineto Romano 42, I-00156 Rome, Italy}

\date{August 27, 2009}

\begin{abstract}

We propose a superconducting phase qubit on the basis of the
radio-frequency SQUID with the screening parameter value $\beta_L
\equiv (2\pi/\Phi_0)LI_c \approx 1$, biased by a half flux quantum
$\Phi_e=\Phi_0/2$. Significant anharmonicity ($> 30\%$) can be achieved
in this system due to the interplay of the cosine Josephson
potential and the parabolic magnetic-energy potential that ultimately
leads to the quartic polynomial shape of the well. The two lowest
eigenstates in this global minimum perfectly suit for the
qubit which is insensitive to the charge variable, biased in the optimal
point and allows an efficient dispersive readout. Moreover, the
transition frequency in this qubit can be
tuned within an appreciable range allowing variable qubit-qubit coupling.

%\verb+\pacs{#1}+ command.

\verb  PACS numbers: 85.25.Dq, 74.50.+r, 03.67.Lx, 03.67.Pp

\end{abstract}
\maketitle

The superconducting qubits based on the Josephson tunnel junctions
(see, e.g., the reviews in Refs. \cite{Makhlin,Devoret2004}) have
already demonstrated their great potential for the quantum
computation \cite{WendinShumeiko}. The so-called phase qubits
present the class of devices which are particularly suitable for
integration with microwave on-chip transmission lines and
resonators, i.e. the elements which significantly extend the scope
of the quantum circuit designs \cite{Martinis2009}. These qubits are based
on the energy quantization in the shallow wells of the inclined cosine
Josephson potential \cite{Martinis}. This shape is ensured either by
finite bias current $I_s$ with the value slightly below the critical
current of the Josephson junction $I_c$ or a finite flux bias $\Phi_e$
applied to the qubit loop (in the case of a loop configuration of
the circuit) \cite{Steffen2003}. In both cases, the energy potential
can be approximated by the cubic parabola with a smooth energy
barrier isolating the well from one side and allowing escape out of this well
enabling a simple readout.

The low depth of the cubic parabola well leads to anharmonicity, viz.
successive reduction of the transition energies $\Delta E_n = (E_{n+1} -
E_{n}),$ $n=0, 1, ...$, from bottom to top, necessary for the qubit
operation within the basis states $|n=0\rangle$ and $|n=1\rangle$,
excluding unwanted excitation of the higher energy states ($n>1$).
Usually the phase qubit is designed such that for appropriate phase
bias the cubic potential well includes three-four energy levels with
anharmonicity of a few per cent \cite{Devoret2004,Steffen2003}. This
is achieved by adjusting the plasma frequency of the Josephson
junction both by designing appropriate parameters of the junction
and, possibly, by applying external capacitor shunting. The lowering
of the energy barrier by applying the so-called measuring pulse,
makes possible the reduction of the number of the
levels to two ($n=0$ and 1), with notably different rates of escape to a running-phase
state (in the case of current bias), or to the lower-energy
state in the adjacent well (in the case of the loop configuration of the
qubit). The large (but finite) difference of these tunneling rates
sets the maximum theoretical value for the fidelity of such measurement
to 96.6\%. In the carefully designed and optimally biased qubit the
best experimental fidelity values approach 90\% \cite{Lucero}. The
main disadvantage of such phase qubit is the necessity of resetting it
after each measurement.

\begin{figure}[b]
\begin{center}
\includegraphics[width=3in]{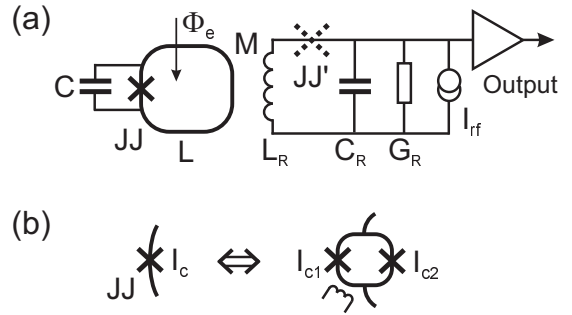}
\caption{(a) Electric diagram of the qubit coupled to a resonant circuit and (b) possible
equivalent compound (two-junction SQUID) circuit of the Josephson element included into the qubit loop.
Capacitance $C$ includes both the self-capacitance of the junction and the external capacitance.
Due to inclusion in the resonant circuit of a Josephson junction JJ', the resonator
may operate in the nonlinear regime, enabling a bifurcation-based readout.} \label{EqvSchm}
\end{center}
\end{figure}

In contrast to the charge \cite{Nakamura}, charge-phase
\cite{Vion,Z-Phys-C}, flux \cite{Chiorescu}, transmon \cite{Koch}
and recently proposed, the so-called fluxonium \cite{Manucharyan-arx09} qubits,
the conventional phase qubits cannot inherently operate in an
optimal point, i.e. in the symmetric working point insensitive in the first
order to the noise that could give drastic improvement to the qubit
performance \cite{Vion,Z-JETP}. (The exceptions are the
recently proposed three-junction interferometer circuit
\cite{Chiarello} and the so-called camelback potential phase qubit
based on the two-junction SQUID \cite{Hoskinson}.) Moreover, the
limited anharmonicity of the phase qubit makes the observable
reactance impedance (i.e. the Josephson inductance) values in the
ground and exited states hardly distinguishable. This poses serious
problems for dispersive readout schemes, which proved advantageous where applicable
\cite{Z-Phys-C,Z-JETP,Lupascu2004} and have allowed quantum nondemolition measurements as well as
high fidelity measurements based on bifurcation amplifiers \cite{Siddiqi2004,Lupascu2006}.

In this paper we propose an improved phase qubit with significant
anharmonicity, in which the manipulation and dispersive readout are
both possible in a symmetry point.
The circuit diagram of our qubit is shown
in Fig.\,1a. It comprises the superconducting loop with geometrical
inductance $L$ closed by a Josephson junction, generally
shunted by an external capacitance, and rf-driven $L_RC_RG_R$
resonance circuit inductively coupled to the qubit loop. The
peculiarity of this qubit is the unity value of the SQUID screening parameter
$\beta_L \equiv (2\pi/\Phi_0)LI_c \approx 1$, where $\Phi_0=h/2e$ is
the flux quantum. This can be achieved
by an accurate design of the circuit including replacement of the
single junction with a two-junction SQUID, allowing more precise
adjustment of the resulting critical current $I_c$ (see Fig.\,1b).

The potential energy of the stand-alone qubit biased by external
magnetic flux $\Phi_e$ includes the magnetic and Josephson
components and can be written as
\begin{equation}
\label{U-phi} U(\phi, \phi_e) = E_L \left[0.5(\phi
-\phi_e)^2-\beta_L (1+\cos \phi) \right],
\end{equation}
where $E_L = (\Phi_0/2\pi)^2/L = E_J/\beta_L$ is characteristic
magnetic energy associated with the loop inductance, the Josephson
coupling energy $E_J=(\Phi_0/2\pi) I_c$, the phase variable $\phi$ and
the phase bias $\phi_e =
2\pi\Phi_e/\Phi_0$. For small
values of $\beta_L \ll 1$, the potential Eq.\,(\ref{U-phi}) yields
the almost parabolic shape of the global single well (first term in
Eq.\,(\ref{U-phi})), whereas for the values $\beta_L$ appreciably
greater that 1, the series of wells are superimposed on the global
parabola, so the bottom parts of these local minima can also be
approximated by the quadratic parabolas. In the case of large density of the levels
within these parabolas, the
energy spectrum is also close to that of a harmonic oscillator.
So, neither of these cases allows significant anharmonicity necessary
for the efficient qubit operation.

\begin{figure}[b]
\begin{center}
\includegraphics[width=3.4in]{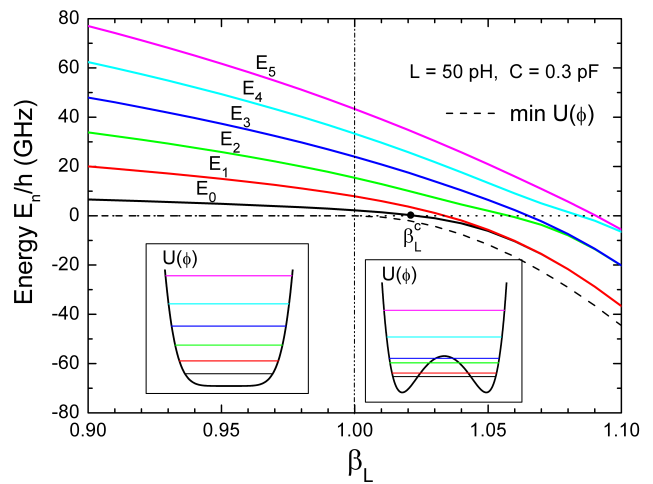}
\caption{(Color online) Position of the lowest six levels (solid
lines) in the potential Eq.\,(\ref{U-phi}) for $\phi_e=\pi$ as a
function of parameter $\beta_L$ for typical values of $L$ and $C$,
yielding $E_J/E_c \sim E_L/E_c \approx 5.1 \times 10^4$. With an
increase of $\beta_L$, the spectrum crosses over from that of the
harmonic oscillator type (left inset) to the set of the doublets (right inset),
corresponding to the weak coupling of the oscillator-type states in two separate
wells. The spectrum in the central region $\beta_L \approx 1$ is
strongly anharmonic. The dashed line shows the bottom energy of the
potential $U(\phi,\phi_e=\pi)$, which in the case of $\beta_L
> 1$ is equal to $-\Delta U\approx -1.5 E_L (\beta_L-1)^2/\beta_L$ (in
other words, $\Delta U$
is the height of the energy barrier in the right inset)
\cite{Chiarello2000,Chiarello2007}. The dotted (zero-level) line indicates the
energy in the symmetry point $\phi = 0$, i.e. at the bottom of the
single well ($\beta_L\leq 1$) or at the top of the energy barrier
($\beta_L > 1$). The black dot shows the critical value $\beta^c_L$
at which the ground state energy level touches the top of the
barrier separating the two wells. } \label{En_b} \label{levels-beta}
\end{center}
\end{figure}

The essentially different shape of the potential Eq.\,(\ref{U-phi})
with $\phi_e=\pi$ is, however, achieved for $\beta_L \approx 1$,
i.e. when the quadratic magnetic term is partially compensated by the
quadratic term in the Josephson energy expansion near the bottom
$(U=0)$ of the single well centered at $\varphi \equiv \phi -\pi =
0$, i.e.
\begin{equation}
\label{U-phi-apprx} U(\varphi) \approx E_J \left[ -\frac{(\beta_L-1)}{2\beta_L}\varphi^2+
\frac{1}{24}\varphi^4+O(\varphi^6)\right].
\end{equation}
In the ultimate case of $\beta_L=1$, Eq.\,(\ref{U-phi-apprx})
provides the obtuse shape of the quartic parabola. Taking into
account the finite kinetic energy of the system in corresponding
Schr\"{o}dinger equation, $\hat{Q}^2/2C = -4E_c
\partial_{\phi\phi}$, where $\hat{Q}=-i(2e) \partial_{\phi}$ is the
charge operator \cite{Anderson}, one can perform quantization of the
system. Application of the quasiclassical quantization rule of
Bohr-Sommerfeld yields for this quartic oscillator the energy levels
obeying the 4/3 power law \cite{Sanchez}:
\begin{equation}
\label{B-S-apprx} E_n^{(\textrm{qc})} = \epsilon (n+1/2)^{4/3},
\end{equation}
with prefactor $\epsilon$ which in terms of the parameters of our
circuit is equal to
\begin{equation}
\label{B-S-apprx2} \epsilon = 2^{-5/3} 3
\,[\pi/\textrm{K}(1/2)]^{4/3} (E_J E_c^2)^{1/3} \approx 1.9 (E_J
E_c^2)^{1/3},
\end{equation}
where $\textrm{K}(k)$ is the complete elliptic integral of the first
kind. Thus, the energy spectrum in the quartic potential takes
intermediate position between the equidistant spectrum of the
harmonic oscillator $E_n \propto (n+1/2)$ and that of the
rectangular well, $E_n \propto (n+1)^2$, having extremely high
anharmonicity. Expressions (\ref{B-S-apprx}) and (\ref{B-S-apprx2}) are
exact for the higher levels ($n\gg 1$) and large "mass" (capacitance
$C$), ensuring the very large ratio of the Josephson energy $E_J$ to the
charging energy $E_c=e^2/2C$. An estimate of the anharmonicity
factor in this quasiclassical approximation can be immediately
obtained from Eq.\,(\ref{B-S-apprx}):
\begin{equation}
\label{delta} \delta_{\textrm{qc}} =(\Delta E_1-\Delta E_0)/\Delta E_0 \approx 26\%.
\end{equation}
The numerical solution of the corresponding Schr\"{o}dinger
equation with potential energy Eq.\,(1)
yields in the limit $E_J/E_c \gg 1$ an even larger value of
the anharmonicity factor, $\delta \approx 33\%$ (see the energy spectrum in Fig.\,2).
These values
substantially exceed the typical anharmonicity values of the
conventional phase qubit, $|\delta_{\textrm{phase}}| \approx\,3\%$,
for the number of levels inside the cubic-parabola well equal to
four \cite{Devoret2004,Steffen2003}, and transmon-qubit,
$|\delta_{\textrm{transmon}}| \approx (E_c/8E_J)^{1/2} \lesssim 5\%$
for optimum values $E_J/E_c \gtrsim 50$ \cite{Koch}. Moreover, in contrast to the
negative values of $\delta$ in these examples, the series of the
energy levels in the quartic potential has positive $\delta > 0$,
i.e. corresponds to successively increasing level spacings $\Delta
E_1 < \Delta E_2 < \Delta E_3 ...$

\begin{figure}[b]
\begin{center}
\includegraphics[width=3.2in]{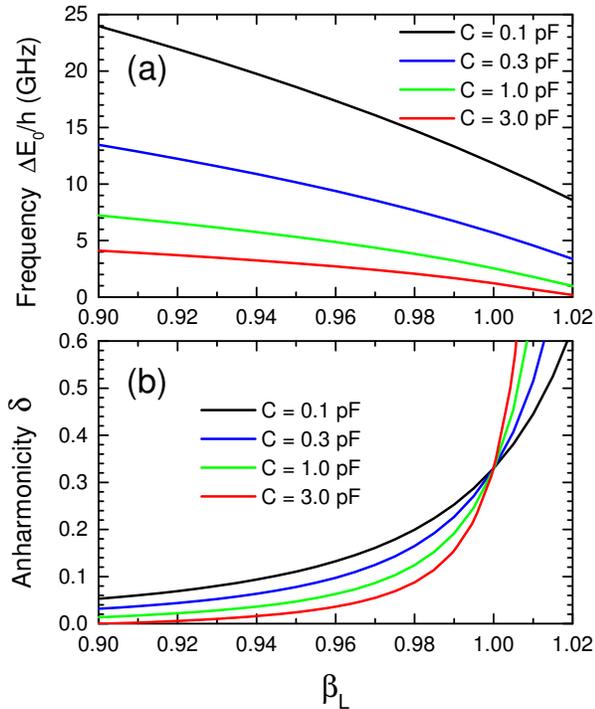}
\caption{(Color online) (a) The qubit frequency as a function of parameter
$\beta_L$ for fixed $L=50$\,pH and several values of capacitance
$C=0.1$, 0.3, 1.0 and 3.0\,pF (from top to bottom), corresponding to the values of the ratio $E_L/E_c\approx
1.7\times 10^4, 5.1\times 10^4, 1.7\times 10^5$ and $5.1\times 10^5$.
(b) Anharmonicity parameter $\delta$ as a
function of parameter $\beta_L$ for the same as in (a) inductance $L$ and capacitance values
(from top to bottom).} \label{frequency-anharmonicity}
\end{center}
\end{figure}

Such a large, positive anharmonicity is a great advantage of the quartic
potential qubit allowing manipulation within the two basis qubit
states $|0\rangle$ and $|1\rangle$ not only when applying resonant microwave field,
$\nu_{\mu\textrm{w}} \approx \nu_{10}$, but also when applying control
microwave signals with large frequency detuning or
using rather wide-spectrum rectangular-pulse control signals. The characteristic qubit
frequency $\nu_{10}=\Delta E_0/h$ and the anharmonicity factor
$\delta$ computed from the Schr\"{o}dinger equation for the original
potential Eq.\,(\ref{U-phi}) in the range $0.9 \leq
\beta_L\leq 1.02$ are shown in Fig.\,\ref{frequency-anharmonicity}. One
can see that the significant range in the tuning of the qubit
frequency within the range of sufficiently large anharmonicity ($\sim
20-50\%$) is attained at a rather fine (typically $\pm1\textrm{-}2\%$) tuning of $\beta_L$
around the value $\beta_L=1$. Such tuning of $\beta_L$ is possible in the
circuit having the compound configuration shown in Fig.\,1b.
For values of $\beta_L>1$, the symmetric energy potential has two minima and a
barrier between them. The position of the ground state level depends on
$\beta_L$ and the ratio of the characteristic energies $E_J/E_c = \beta_L E_L/E_c$.
The value of $\beta_L$ at which the ground state level touches the top of
the barrier sets the upper limit $\beta^c_L$ for the quartic qubit
(marked in Fig.\,\ref{levels-beta} by solid dot).
At $\beta_L >\beta^c_L$, the
qubit energy dramatically decreases and the qubit states are nearly the
symmetric and antisymmetric combinations of the states inside the
two wells (see the right inset in Fig.\,2). Although the qubit with such parameters has very large anharmonicity
and can be nicely controlled by dc flux pulses \cite{Chiarello2007,Poletto},
its readout can hardly be accomplished in a dispersive fashion.

\begin{figure}[b]
\begin{center}
\includegraphics[width=3.3in]{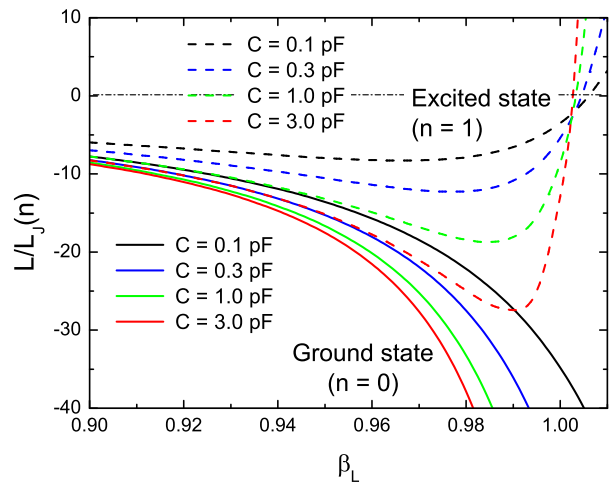}
\caption{(Color online) The values of the Josephson inductance of
the quartic potential qubit in the ground (solid lines) and excited
(dashed lines) states calculated for the geometric inductance value $L=50$\,pH
and the set of capacitances $C$, increasing from top to bottom for
both groups of curves.} \label{L_J} \label{inductance-L01}
\end{center}
\end{figure}

Another advantage of the phase qubit having the energy potential of the shape close to the
quartic one is a strong dependence of its Josephson inductance $L_J(\Phi_e,n)$ on the
quantum state $|n\rangle$. The observed value of the reverse inductance is
related to the local curvature of the dependence of corresponding energy $E_n$
on flux $\Phi_e$ (see, e.g., Ref.\cite{Z-JETP}),
\begin{equation} \label{L_J} L_{J}^{-1}(\Phi_e,n) =
\frac{2\pi}{\Phi_0}\langle n| \frac{\partial \hat{I}(\phi,\phi_e)}{\partial \phi_e} |n\rangle
= \frac{\partial^2 E_n(\Phi_e)}{\partial \Phi_e^2},
\end{equation}
where $\hat{I}$ is the operator of supercurrent circulating in the qubit loop.
The dependence of the reverse inductance $L_{J}(\Phi_e=\Phi_0/2,n)$
calculated numerically in the two lowest quantum states ($n=0$ and 1) for $L=50$\,pH
and the same set of capacitances $C$ as in Fig.\,\ref{frequency-anharmonicity}
is shown in Fig.\,\ref{inductance-L01}.
One can see that the ratio of the geometrical to Josephson inductances $L/L_J$ takes large and very different
values that can be favorably used for the dispersive readout, ensuring a sufficiently large output signal.
Note that for $\beta_L<1$, both inductances $L_J(n=0)$ and $L_J(n=1)$ are negative,
whereas at $\beta_L>1$ the inductance $L_J(n=1)$ changes the sign to positive.

\begin{figure}[t]
\begin{center}
\includegraphics[width=3.2in]{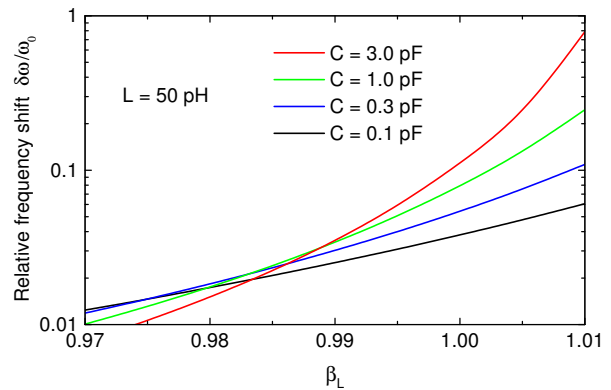}
\caption{(Color online) The resonance frequency shift in the circuit
due to excitation of the qubit with the inductance value $L=50$\,pH
and the set of capacitances $C$, decreasing from top to bottom. The dimensionless coupling
coefficient $\kappa=0.05$.} \label{dOmega} \label{f-shift}
\end{center}
\end{figure}

The readout of this qubit is based on the measurements of the
reactive part (inductance) of the loop impedance probed by a low-frequency
ac signal, $f \ll \nu_{10}$ of sufficiently small amplitude \cite{Rifk} (see also Ref. \cite{Z-Phys-C}).
This signal is supplied by a rf-driven oscillator (Fig.\,1a) as an
alternating biasing flux, $\Phi_e = 0.5\Phi_0 + M I_L $ or $\phi_e = \pi + \delta\phi_e$,
where $\delta\phi_e = a \cos(2\pi ft)$ with
$a\ll1$. Here $M=\kappa (LL_R)^{1/2}$ is mutual inductance, $\kappa$, a dimensionless
coupling coefficient and $I_L$, the ac current in inductance $L_R$. Coupling of the qubit
to the resonance tank circuit causes renormalization of the circuit inductance
(see, e.g. \cite{Z-Phys-C,Z-JETP}),
\begin{equation} \label{frequency-shift} L_{R}^{(n)}=
L_R(1-\kappa^2 L/L_J(n)),
\end{equation}
and the resonance frequencies $\omega_n=[L_R^{(n)}C]^{-1/2}$, where $n = 0$ and 1. The relative
difference of the resonance frequencies for the qubit in the excited and
ground states is
\begin{equation} \label{frequency-shift}
\frac{\delta\omega}{\omega_0}= \frac{\omega_1-\omega_0}{\omega_0} =
\sqrt{ \frac{1-\kappa^2 L/L_J(0)}{1-\kappa^2 L/L_J(1)} } -1.
\end{equation}
Figure \ref{f-shift} shows this relative frequency shift versus parameter
$\beta_L$. One can see
that for the rather conservative value of dimensionless coupling
$\kappa=0.05$, the relative frequency shift can achieve the easily measured values of about $10\%$.
The efficiency of the dispersive
readout can be improved in the non-linear regime with bifurcation \cite{Siddiqi2004}.
With our device this regime can be achieved in the resonance circuit
including, for example, a Josephson junction (marked in the diagram in Fig.\,1 by a dashed cross).
Due to the high sensitivity of the amplitude (phase) bifurcation to the
threshold determined by the effective resonance frequency of the circuit, one can expect a
readout with high fidelity even at a rather weak coupling of the
qubit and the resonator (compare with the readout
of quantronium in Ref.\,\cite{Siddiqi2006}). Further improvement of the readout
can be achieved in the QED-based circuit including this qubit \cite{Metcalfe}.

The loop configuration and frequency detuning of the quartic qubits
should allow their inductive coupling with variable strength keeping both qubits in optimal
points. Variable coupling of the optimally biased qubit to a superconducting resonator is also possible.
More sophisticated coupling of the pairs of quartic qubits can be accomplished,
for example, using a Josephson-junction coupler in a fashion recently proposed
by Harris et al. \cite{Harris2009}.

In conclusion, we have shown that the phase qubit of the rf-SQUID
configuration with parameter $\beta_L \approx 1$ and flux bias
$\Phi_e=\Phi_0/2$ has remarkable characteristics. Still, we expect
that implementation of this qubit requires the solution of several
experimental problems. For example, due to a high sensitivity of the
qubit parameters to the magnitude of $\beta_L$, whose optimum values
lie within a rather narrow range ($\pm 1\textrm{-}2\%$), particular
precaution should be taken against fluctuations in the line
controlling the effective Josephson coupling in the circuit (see
Fig.\,1b), because otherwise it may cause significant dephasing of the qubit.
Furthermore, flux bias $\Phi_e=\Phi_0/2$ should also be set as precisely as
possible. Experimentally, it can be realized either by freezing the
flux $\Phi_e=\Phi_0$ in the main loop having a symmetric gradiometer configuration
\cite{Majer}, or by including in the loop of a Josephson
$\pi$-junction \cite{Ioffe99} with a sufficiently high critical
current ensuring the steady phase shift of $\pi$.

Of course, similar to properties of the conventional types of the phase qubits,
the coherence characteristics of the quartic qubit will be strongly
dependent on the material properties of the circuit. Minimizing the losses
due to the qubit coupling to microscopic degrees of freedom (two-level systems) inside
the dielectrics surrounding the superconducting circuit (the substrate,
insulator inside the capacitor, the junction barriers, etc.) play crucial
role for improving the qubit coherence \cite{Martinis2009}.
Since the operation and tuning of the quartic
qubit is possible without leaving the optimal point, one may expect a
weaker coupling of the qubit to these microscopic two-level
systems located inside dielectrics and, therefore, a better quantum coherence.
Moreover, the zero persistent supercurrent circulating in the qubit loop at
the optimum bias, $\phi_e =\pi$, may also reduce the effect of quasiparticle
tunneling on the qubit coherence.
Probably, such weakening of the qubit coupling to external degrees of freedom can
explain reasonably good coherence characteristics ($\tau_{\textrm{Rabi}}
\approx 60$\,ns) of the Nb camelback qubit operated in an optimal point
at zero persistent current \cite{Hoskinson}. Anyway, the properties of the proposed
quartic qubit will be clarified in experiment which is currently
in the preparation stage.

We thank Michael Wulf and Ralf Dolata for useful discussions.
This work was partially supported by the EU through the EuroSQIP
project and DFG (German Science Foundation) through grant ZO124/2-1.

\end{document}